# Colloidal Flying Carpets

Nienke Geerts[1] and Erika Eiser[2]

1.  FOM Institute for Atomic and Molecular Physics [AMOLF], Science Park 113, 1098XG Amsterdam, The Netherlands.
2.  Cavendish Laboratory, University of Cambridge, J.J. Thomson Avenue, Cambridge CB3 0HE, and BP Institute, Cambridge CB3 0EZ, United Kingdom.

**DNA plays a special role in polymer science not just because of the highly selective recognition of complementary single DNA strands but also because bacteria can express DNA chains that are very long yet perfectly monodisperse. The latter reason makes long DNA molecules widely used as model systems in polymer science. Here, we report the unusual self-assembly that takes place in systems of colloids coated with very long double-stranded DNA. In particular, we find that colloids coated with such long DNA can assemble into unique "floating" crystalline monolayers. Floating colloidal structures have potentially interesting applications as such ordered structures can be assembled in one location and then deposited somewhere else. This would open the way to the assembly of multi-component, layered colloidal crystals.**

Much of the interest in DNA-coated colloids derives from the fact that such building blocks can be used to make complex, self-assembling materials, because of the high selectivity of hybridization of complementary single-stranded (ss)DNA sequences [1-16]. However, DNA-coated colloids are also interesting because DNA can be made very long, yet monodisperse [17]. In the present Letter we report on the unusual self-assembly properties of colloids coated with a single type of long, double-stranded DNA.

Colloids coated with such DNA polymers have the ability to form "floating" crystalline monolayers. This behavior is not found in colloids coated with shorter chains. In our experiments, we used colloids coated with DNA that is either blunt or terminated with a single-stranded sequence that can selectively bind to the functionalized surface of the sample cell. However, the colloids cannot bind to each other via DNA hybridization. As we shall show below, the presence or absence of sticky (ssDNA) ends is only of importance for colloids coated with short DNA sequences. The sticky ends are not relevant for the formation of floating colloidal monolayers as these structures only form when the colloids are coated with long DNA.

As a sample chamber we use a 96-well plate (Sensoplate; Greiner bio-one), allowing us to run many experiments under identical conditions. Onto the bottom glass surfaces of these wells we grafted a polymer monolayer holding ssDNA "sticky ends" [Fig. 1(a)]. The surface coverage was tested with colloids coated with the complementary 12 bases [Fig. 1(b)]. The polymer monolayer always prevents a-specific binding of colloids without DNA to the surface [Fig. 1 (a) and (c)]. For our experiments we use polystyrene (PS) particles (1 µm diameter; Invitrogen) coated with "long", "intermediate" and "short" DNA sequences. The "long" DNA chains were obtained from bacteriophage lambda DNA (λ-DNA; New England Biolabs), which is monodisperse and has a 16 µm contour length (485000 bps; radius of gyration ~ 800 nm [16]). The λ-DNA, predominantly circular at room temperature, can be linearised by heating. The two ends of the linear dsDNA are the same complementary 12-base single strands as those used for testing the surface coverage. One of these ends was first hybridized to an oligonucleotide bound to biotin. After hybridization the backbone was ligated to prevent later dissociation. As described elsewhere [8, 16] the biotin ends are then used to attach the DNA to the PS-colloids. The resulting beads carry long dsDNA spacers with ssDNA "sticky ends" that can bind (hybridize) to the complementary ssDNA on the surface. Hybridization between the particles is not possible as they all display the same ssDNA sequence (5'-AGGTCGCCGCCC-3'). As the radius of gyration ($R_g$) of λ-DNA is similar to the particle-size used, no more than 8 to10 strands can bind to the colloids [16]. The intermediate length DNA was derived from a different plasmid that contains the same 12 bp overhangs (pBeloBac11; New England Biolabs, ER2420S). This plasmid comprises only 7500 bps and the resulting DNA chain has a radius of gyration of about 200 nm that is somewhat smaller than the radius of the colloids and therefore up to ~25 DNA coils can bind per particle [8]. Finally, a third set of DNA-coated colloids was functionalized with the short, 12-nucleotide ssDNA strand that can bind to the complementary ssDNA on the surface of the sample cell. In this case a few thousand short nucleotides can bind per particle. Sedimentation was minimized by density matching the DNA-coated PS-beads with a sucrose solution of 150 mg/ml in a TRIS-HCl buffer (100 mM; pH 8). In all experiments we worked at a low colloidal volume fraction of $\phi \approx 4.10^{-4}$.

As we use fluorescently labeled PS-colloids (Invitrogen; red 580/605) we can follow the colloidal aggregation in time by confocal microscopy (DMIRB, Leica; spinning disc scan head (CSU22, Yokogawa Electric Corp.; 60x water immersion objective). Two hours after injecting a solution of colloids coated with λ-DNA into the sample cell, we observe the appearance of single layers of close-packed colloids. Interestingly, these colloidal sheets float above the bottom of the cell at heights ranging from the $R_g$ of λ-DNA up to 5 µm (S1). If left at room temperature for longer times, these structures can grow into large colloidal carpets that span the entire field of view of the microscope [Fig. 2(a)]. The structures are clearly crystalline, be it that some crystals contain point defects or even grain boundaries. In fact, the 2D crystals can even show clear facets [Fig. 3(a)-3(b)] indicating that the crystallites have some time to anneal during growth. Indeed, the pair-correlation function of this 'colloidal flying carpet' displays distinct peaks corresponding to those of a 2D hexagonal crystal [see red arrowheads, Fig. 2(b)]. The crystals remain intact and in place when the sample is reheated to a temperature where the DNA links would be broken [Fig 4(a) – 4(c)]. This indicates that DNA hybridization plays no role in the stabilization of the 2D colloidal crystals. We observe the formation of these 'flexible' colloidal carpets only above the bottom surface of the cell, but not in the bulk of the solution.

The experiments suggest that few of the ssDNA on the colloids are attached to the surface. This follows from the following observation: if all strands of DNA would anchor to the surface the carpets would be barely mobile. However, this seems not to be the case as the carpets can move over appreciable distances. In Figure 3(a) we show a carpet with a low-symmetry shape at a given time. After leaving the sample for two hours we imaged the same carpet again. There was no sign of further growth, but the carpet did rotate [Fig. 3(b)], indicating that either not all strands of DNA were hybridized or rearrangements are possible at room temperature (well below the melting temperature of our 12 base pair bonds).

When we repeated the above experiments with the intermediate-length pBelo-DNA (7500 bps, $R_g$ ~200 nm) we again observed the formation of floating 2D crystal structures (data not shown). In this case the maximum sizes of the colloidal sheets seemed smaller than for the λ-DNA coated beads. A recent paper on DNA adsorption on PLL-PEG films [18] inspired us to repeat the experiment without ssDNA on our surface or on our beads. The results indicated that DNA hybridization is not necessary to obtain the crystal structures that we observe. However, when we repeated the experiments with colloids coated with "short" DNA (no dsDNA spacer; 12 bases ssDNA attached directly to the colloids), no crystalline sheets are formed. Rather, we observe an amorphous colloidal layer in direct contact with the polymer layer coating the surface of the sample well [Fig. 4 (d) and 4(e)]. This randomness as well as the fact that these colloids can be 'melted off' above the hybridization temperature of the 12bp-bonds [Tm ≈ 45 °C; Fig 4(f)] suggests that two factors are essential for the formation of crystalline colloidal membranes: 1) weak binding to the surface that allows colloids to diffuse and 2) weak steric stabilization of the colloids against the formation of direct contacts that are stabilized by dispersion forces. If the colloids are too strongly bound to the surface of the cell, as is the case for the colloids coated with many short ssDNA strands, surface diffusion of the colloids is inhibited and hence crystalline layers can neither form nor anneal. In addition, the short DNAs provide steric stabilization against the formation of direct colloid-colloid contacts. Hence, neither condition for the formation of crystalline colloidal carpets is satisfied in the case of colloids coated with short ssDNA.

Previously we have shown that colloids grafted with either λ-DNA, or pBelo-DNA with the same ssDNA overhangs do not aggregate in the bulk under the conditions used in the present experiments [8]. This implies that in the bulk of the solution, where the colloidal concentration is low, the DNA cloud that surrounds the colloids is sufficient to prevent aggregation due to non-specific Van der Waals interactions. However, near the bottom cell surface, this situation changes: the charged DNA strands on the colloids are attracted weakly and non-specifically to the (oppositely charged) poly-lysine layer on the surface. This weak attraction could also be observed in experiments where unbound fluorescent λ−DNA (in the absence of bare colloids) was found to absorb weakly to the polymer-coated surface of the sample cell. As a consequence of the DNA-surface attraction, the colloidal concentration is significantly enhanced on the surface and colloids come into contact sufficiently frequently to overcome the weak entropic stabilization provided by the long dsDNA. Indeed, in the 2D crystals the colloids are effectively touching each other and the thermal stability in the colloidal membranes is therefore most likely due to the action of short-range dispersion forces. At the same time, the fact that highly ordered 2D crystals form indicates that crystal growth is slow: aggregation is definitely not diffusion limited. This crystallization mechanism is very different from previous observations where 2D crystallization from a very dilute colloid/polymer mixture is observed (see review [19]). In this case a depletion attraction between colloids and the support surface is induced by the presence of non-adsorbing polymers leading to small 2D crystallites that are firmly pressed onto the support surface. Repeating our experiments under similar conditions as described above, but in which we do not bind λ−DNA to the colloids we see random aggregation in the bulk and no crystallization at the surface [SI 2].

We also investigated the possibility that sucrose, which we used for density matching, may have an influence on the 2D-aggregation. We therefore used another density-matching solvent where the TRIS-HCl buffer is mixed with $D_2O$ (50-50; 100 mM final concentration). Also in these conditions we obtain colloidal flying carpets, indicating that sucrose plays no key role in the formation of flying carpets (data not shown).

To summarize, we found a new way to crystallize DNA-coated colloids in a loosely bound 2D carpet hovering over the support surface. The crystallization mechanism seems to rely on weak non-specific adsorption of the DNA-coated colloids (or, more precisely, of the DNA coating of these colloids) to the bottom of the sample cell [18]. Under these conditions, the steric stabilization of the colloids by the grafted dsDNA is insufficient to prevent the slow formation of dense structures stabilized by short-ranged dispersion forces. Colloids that are coated with very short ssDNA strands bind strongly to the surface via specific interactions between the complementary strands and form an amorphous adsorbate rather than a crystalline floating carpet.

The ability to make floating, yet surface-bound structures, could provide an interesting route to make novel colloidal structures: in particular, by making use of suitable DNA linkers, it should be possible to induce the self-assembly of multiple layers of 2D crystals that can be attached on top of the carpets. Another option would be to use the free "sticky ends" to close the carpets, forming a tube of colloids with DNA on both sides, as possible docking sides to bind other molecules or materials.

ACKNOWLEDGEMENTS

We thank M. Dogterom for providing laboratory facilities. We are also grateful for useful discussions with D. Frenkel. This work is part of the research program of the Stichting voor Fundamenteel Onderzoek der Materie (FOM), which is supported by the Nederlandse Organisatie voor Wetenschappelijk Onderzoek (NWO).

**Figures**

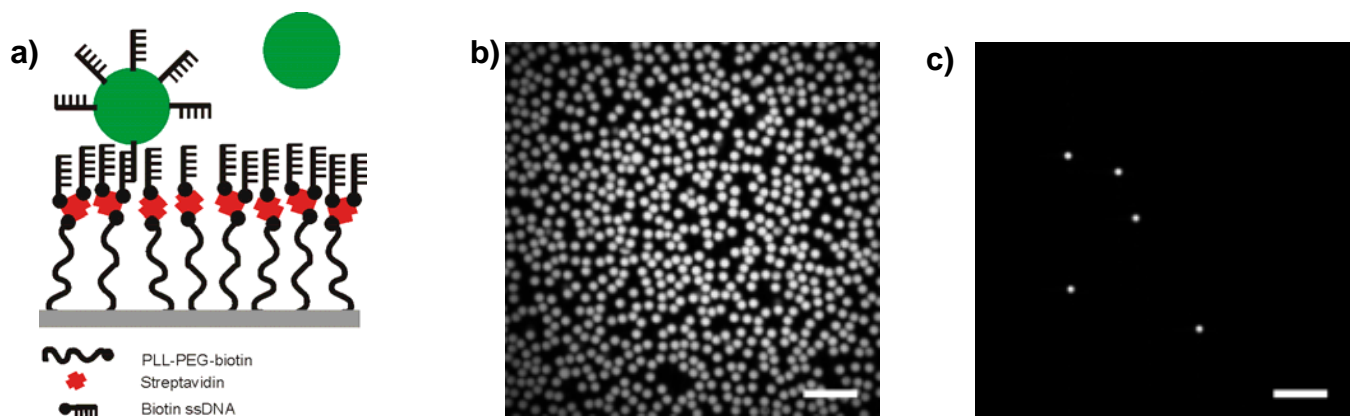

FIG. 1. The DNA coated surface is tested for surface coverage and prevention of a-specific binding. (a) Cartoon of DNA coated surface. Glass is subsequently coated with PLL-PEG-biotin, streptavidin and biotin-ssDNA. Colloids coated with complementary DNA can hybridize to the surface. 'Bare' PS-colloids remain in solution. Images are not to scale. (b) The coverage of a DNA surface can easily be tested with complementary DNA coated beads. (c) a-Specific binding is tested with colloids without DNA coating. Both scale bars 10 μm.

**a)**

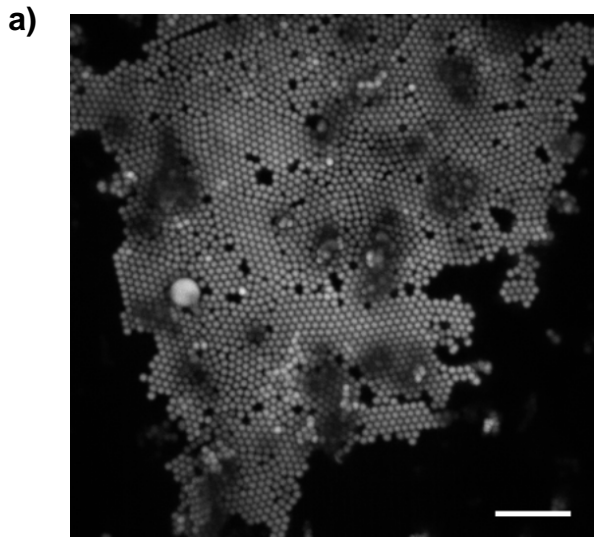

**b)**

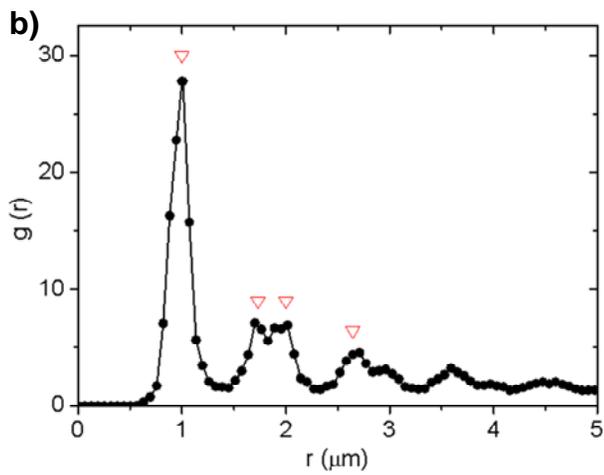

FIG. 2. Aggregation of λ-DNA coated colloids above a DNA coated surface leads to 2D flying carpets. (a) Image of a flying carpet taken with a confocal microscope. As the structure is slightly bent in x and y a z-projection is shown. Scale bar is 10 µm. (b) Pair correlation function of (a). The first neighbor is at contact, the function also shows peaks for the second, third and forth neighbor. ▽ Guide for the eye: positions of expected peaks for a hexagonal crystal.

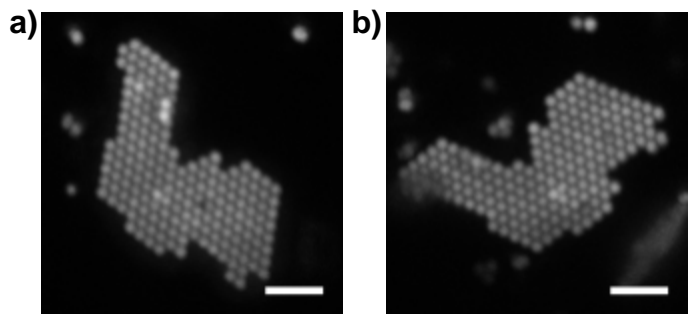

FIG. 3. Flying colloidal carpets are not stuck to the surface; they can move in time. (a) Flying colloidal carpet image taken with confocal microscopy at t = 0 hr. (b) Image of the same carpet taken at t = 2 hrs. Both scale bars 5 μm.

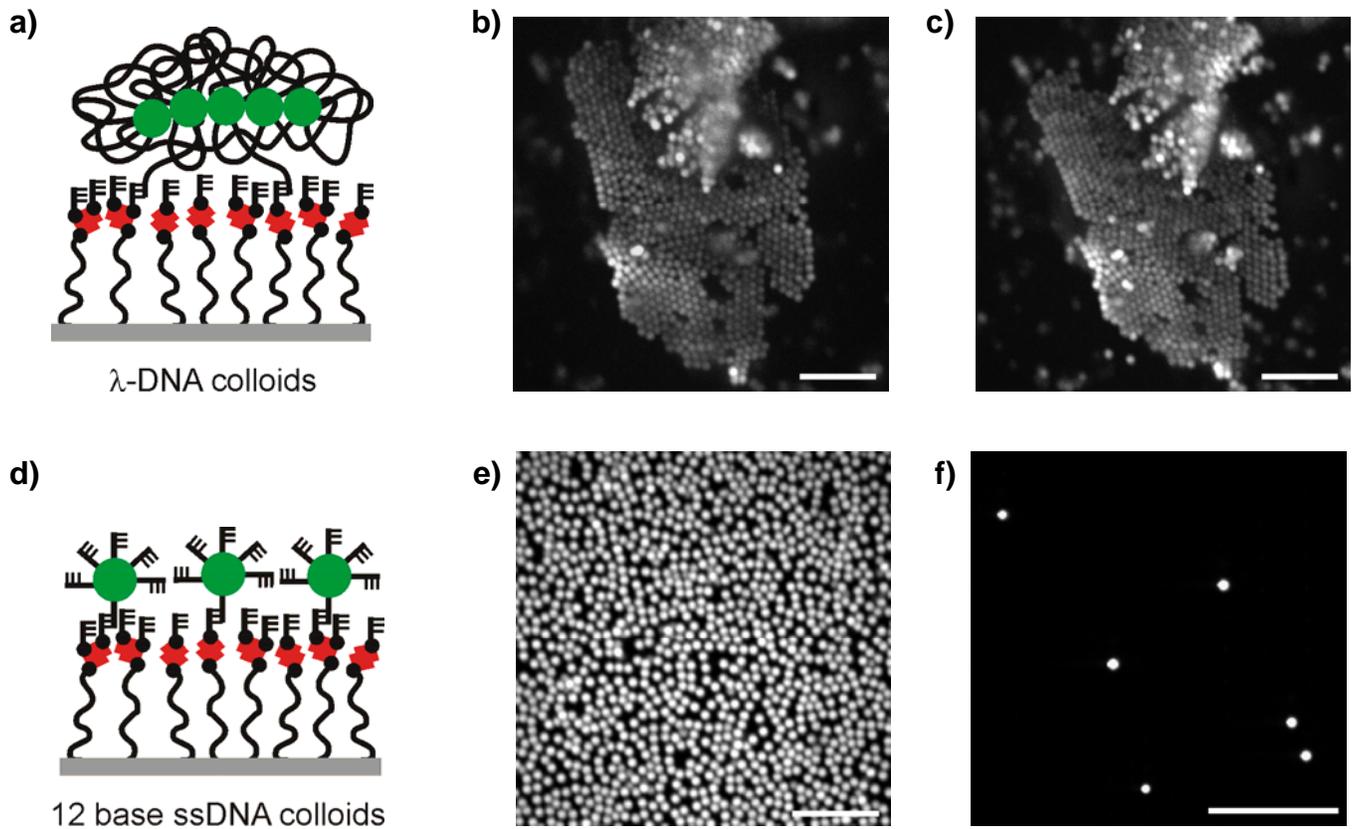

FIG. 4. Short and long DNA-coated colloids respond differently on high temperatures. (a) Schematic cartoon of colloids 'bound' to the surface via λ-DNA that has been attached to the colloids. (b) The resulting 2D carpet formed at RT. (c) Heating the carpet for 7 hours at 70 °C does not melt the crystal. (d) Schematic illustration of colloids bound to the PLL-PEG biotin layer via two complementary 12 base ssDNA sequences, one being attached to the colloids. (e) Typical bound, amorphous layer of such a system formed at room temperature. (f) After heating (70 °C) only few particles remain. All scale bars 10 μm

**Supplementary information**

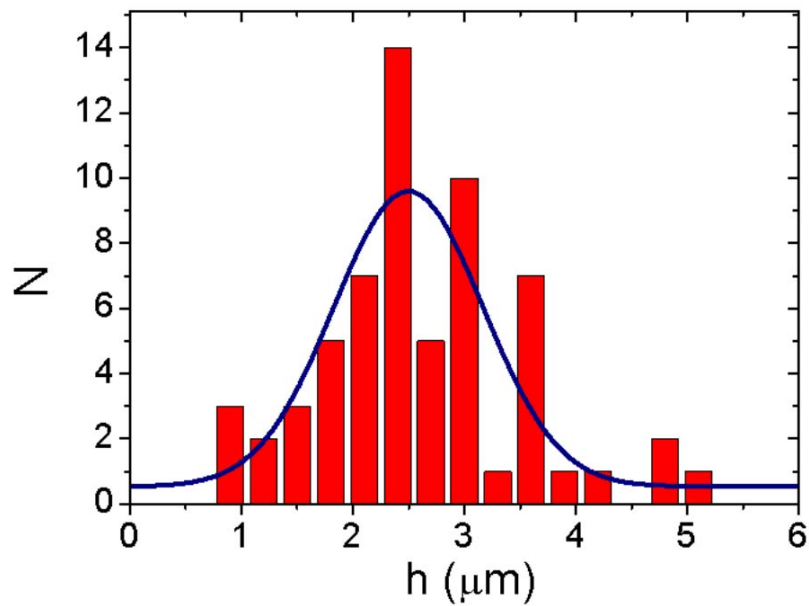

**Fig. S1.** Average flying height of colloidal carpets above the surface. The 2D colloidal crystals lay not directly on the surface. Instead they fly above the surface with heights ranging from λ-DNA $R_g$ up to 5 μm. The bars are real data obtained from three different experiments. The solid line is a Gaussian fit with a peak at an average height of 2.5 μm.

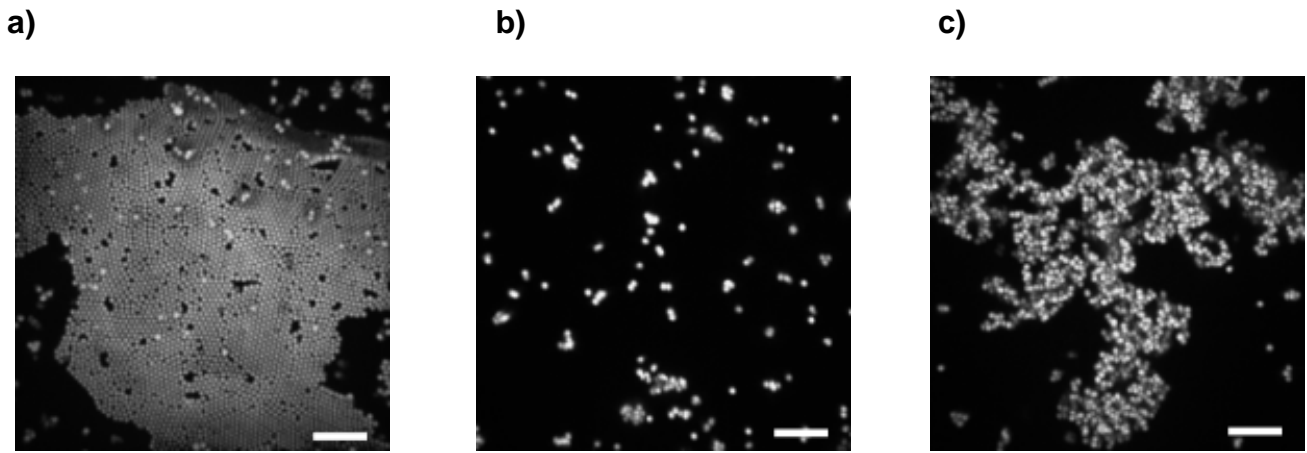

**Fig. S2.** Comparison between aggregation of DNA coated colloids and colloids surrounded by DNA in solution. (a) λ-DNA coated colloids aggregate into a flying colloidal carpet. (b) Colloids surrounded by a solution of Lambda DNA (λ-DNA is present at the same concentration as used in (a)) aggregate hardly; only small 3D clusters appear. (c) If the concentration of λ-DNA in increased 100x, more aggregation is visible. In contrast to (a) this leads to 3D branches rather then 2D carpets. All scale bars are 10 μm.